\documentclass[10pt]{iopart}
\usepackage{graphicx} 

\newcommand{\be}{\begin{equation}}
\newcommand{\ee}{\end{equation}}
\newcommand{\bea}{\begin{eqnarray}}
\newcommand{\eea}{\end{eqnarray}}

\newcommand{\pa}{\partial}

\newcommand{\F}{{\mathcal F}}
\newcommand{\Fo}{{\mathcal F}_o}

\def\tav#1{\left\langle#1\right\rangle}

\begin{document}

\title[All-sky upper limit]
{All-sky upper limit for gravitational radiation from spinning neutron stars}

\author{
P~Astone\dag,
D~Babusci\ddag,
M~Bassan\S,
K~M~Borkowski\#,
E~Coccia\S,
S~D'Antonio\S,
V~Fafone\ddag,
G~Giordano\ddag,
P~Jaranowski\pounds,
A~Kr\'olak\$\footnote[7]{On leave of absence from Institute of Mathematics
Polish Academy of Sciences, Warsaw, Poland.}\footnote[8]{krolak@jpl.nasa.gov},
A~Marini\ddag,
Y~Minenkov\S,
I~Modena\S,
G~Modestino\ddag,
A~Moleti\S,
G~V~Pallottino$\|$,
M~Pietka\pounds,
G~Pizzella\P,
L~Quintieri\ddag,
A~Rocchi\S,
F~Ronga\ddag,
R~Terenzi$^+$ and
M~Visco$^+$}

\address{\dag\ Istituto Nazionale di Fisica Nucleare INFN, Rome, Italy}
\address{\ddag\ Istituto Nazionale di Fisica Nucleare INFN, Frascati, Italy}
\address{\S\ University of Rome ``Tor Vergata" and INFN, Rome II, Italy}
\address{$\|$\ University of Rome ``La Sapienza" and INFN, Rome, Italy}
\address{\P\ University of Rome ``Tor Vergata" and INFN, Frascati, Italy}
\address{$^+$\ IFSI-CNR and INFN, Rome}
\address{\# Centre for Astronomy, Nicolaus Copernicus University,
Toru\'n, Poland}             
\address{\pounds\ Institute of Theoretical Physics, University of Bia{\l}ystok,
Bia{\l}ystok, Poland}
\address{\$\ Jet Propulsion Laboratory, Pasadena, USA}              

\begin{abstract}
We present results of the all-sky search for gravitational-wave 
signals from spinning neutron stars in the data of the EXPLORER resonant bar 
detector. Our data analysis technique was based on the maximum likelihood 
detection method. We briefly describe the theoretical methods that we used 
in our search. The main result of our analysis is an upper limit
of ${\bf 2\times10^{-23}}$ for the dimensionless amplitude of the
continuous gravitational-wave signals coming from any direction in the sky
and in the narrow frequency band from $921.00$ Hz to $921.76$ Hz.  
\end{abstract}


\submitto{\CQG}

\maketitle

\section{Introduction}

A unique property of the gravitational-wave detectors is that with a single
observation resulting in one time series one can search for signals coming from
all sky directions.  In the case of other instruments like optical and radio
telescopes, to cover the whole or even part of the sky requires a large amount
of expensive telescope time as each sky position needs to be observed
independently.  The difficulty in the search for gravitational-wave signals is
that they are very weak and they are deeply buried in the noise of the detector.
Consequently the detection of these signals and interpretation of data analysis
results is a delicate task.  In this paper we present results of an all-sky
search for continuous sources of gravitational radiation.  A prime example of
such a source is a spinning neutron star.  A signal form such a source has
definite characteristics that make it suitable for application of the optimal
detection techniques based on matched filtering.  Moreover such signals are
stable as a result of the stability of the rotation of the neutron star and they
will be present in the data for time periods much longer than the observational
interval.  This enables a reliable verification of the potential candidates by
repeating the observations both by the same detector and by different detectors.
To perform our all-sky search we have used the data of the EXPLORER \cite{exp}
resonant bar detector.  The directed search of the galactic center with the
EXPLORER detector has already been carried out and an upper limit for the
amplitude of the gravitational waves has been established \cite{ROGc00}.

Our paper is divided into two parts.  In the first part we summarize the
theoretical tools that we use in our analysis and in the second part we present
results of our all-sky search.  The main result of our analysis is an upper
limit of $2\times10^{-23}$ for dimensionless amplitude of gravitational waves
originating from continuous sources located in any position in the sky and in
the frequency band from $921.00$ Hz to $921.76$ Hz.

The data analysis was performed by a team consisting of Pia Astone, Kaz
Borkowski, Piotr Jaranowski, Andrzej Kr\'olak and Maciej Pietka and was carried
out on the basis of Memorandum of Understanding between the ROG collaboration
and Institute of Mathematics of Polish Academy of Sciences.  More details about
the search can be found on the website:  {\tt
http://www.astro.uni.torun.pl/\verb!~!kb/AllSky/AllSky.html}.

\section{Data analysis methods}
\label{Sec:Met}

In this section we give a summary of data analysis techniques that we used in 
the search. The full exposition of our method is given in  Ref.\ \cite{puls4}.

\subsection{Response of a bar detector to a continuous gravitational-wave 
signal}
\label{sSec:Res}

Dimensionless noise-free response function $h$ of a resonant bar
gravitational-wave detector to a weak plane gravitational wave in
the long wavelength approximation [i.e.,\ when the size of the
detector is much smaller than the reduced wavelength
$\lambda/(2\pi)$ of the wave] can be written as a linear
combination of the {\em wave polarization functions} $h_+$ and
$h_\times$: 
\be \label{h} h(t) = F_+(t)\,h_+(t) +
F_{\times}(t)\,h_{\times}(t), 
\ee 
where $F_+$ and $F_\times$ are called the {\em beam-pattern functions}
and can be written as 
\bea
\label{bpb} F_+(t) &=& a(t)\cos2\psi+b(t)\sin2\psi, \nonumber
\\[2ex]
F_\times(t) &=& b(t)\cos2\psi-a(t)\sin2\psi,
\eea
where $\psi$ is the polarization angle of the wave and
\bea
\label{abb}
a(t) &=&
\frac{1}{2} (\cos^2\gamma-\sin^2\gamma\sin^2\phi) (1+\sin^2\delta)
\cos[2(\alpha-\phi_r-\Omega_r t)]
\nonumber\\[2ex]&&
+ \frac{1}{2} \sin2\gamma \sin\phi (1+\sin^2\delta)
\sin[2(\alpha-\phi_r-\Omega_r t)]
\nonumber\\[2ex]&&
- \frac{1}{2} \sin^2\gamma \sin2\phi \sin2\delta
\cos(\alpha-\phi_r-\Omega_r t)
\nonumber\\[2ex]&&
+ \frac{1}{2} \sin2\gamma \cos\phi \sin2\delta \sin(\alpha-\phi_r-\Omega_r t)
\nonumber\\[2ex]&&
+ \frac{1}{2} (1-3 \sin^2\gamma \cos^2\phi) \cos^2\delta, \nonumber
\\[2ex]
b(t) &=&
-\sin2\gamma \sin\phi \sin\delta \cos[2(\alpha-\phi_r-\Omega_r t)]
\nonumber\\[2ex]&&
+ (\cos^2\gamma-\sin^2\gamma\sin^2\phi) \sin\delta
\sin[2(\alpha-\phi_r-\Omega_r t)]
\nonumber\\[2ex]&&
- \sin2\gamma \cos\phi \cos\delta \cos(\alpha-\phi_r-\Omega_r t)
\nonumber\\[2ex]&&
- \sin^2\gamma \sin2\phi \cos\delta \sin(\alpha-\phi_r-\Omega_r t).
\eea
In Eqs.\ (\ref{bpb}) and (\ref{abb}) the angles $\alpha$ and $\delta$ are
respectively right ascension and declination of the gravitational-wave source.
The geodetic latitude of the detector's site is denoted by $\phi$,
the angle $\gamma$ determines the orientation of the
bar detector with respect to local geographical directions: $\gamma$ is
measured counter-clockwise from East to the bar's axis of symmetry.
The rotational angular velocity of the Earth is denoted by $\Omega_r$,
and $\phi_r$ is a deterministic phase which defines the
position of the Earth in its diurnal motion at $t=0$ (the sum $\phi_r+\Omega_r
t$ essentially coincides with the local sidereal time of the detector's site,
i.e., with the angle between the local meridian and the vernal point).

We are interested in a continuous gravitational wave, which is described by the
wave polarization functions of the form
\be
\label{wavepol}
     h_+(t) = h_{0+}      \cos\Psi(t), \hspace{5mm} 
h_\times(t) = h_{0\times} \sin\Psi(t), 
\ee 
where 
$h_{0+}$ and $h_{0\times}$ are independent constant amplitudes of the two wave
polarizations. These amplitudes depend on the physical mechanisms
generating the gravitational wave. In the case of a wave
originating from a spinning neutron star these amplitudes can be
estimated by \be \label{hon} h_o =
4.23\times10^{-23}\left(\frac{\epsilon}{10^{-5}}\right)
\left(\frac{I}{10^{45}~\mbox{g cm}^2}\right)
\left(\frac{\mbox{1~kpc}}{r}\right)\left(\frac{f}{\rm
1~kHz}\right)^2, \ee where $I$ is the neutron star moment of
inertia with respect to the rotation axis, $\epsilon$ is the
poloidal ellipticity of the star, and $r$ is the distance to the
star.  The value of $10^{-5}$ of the parameter $\epsilon$ in the
above estimate should be treated as an upper bound.  In reality it
may be several orders of magnitude less.

The phase $\Psi$ of the wave is given by
\bea 
\label{phaza} 
\Psi(t) &=& \Phi_0 + \Phi(t), \nonumber
\\[2ex]
\Phi(t) &=& \sum_{k=0}^{s_1} \omega_k
\frac{t^{k+1}}{(k+1)!} + \frac{{\bf n}_0\cdot{\bf r}_{\rm
SSB}(t)}{c} \sum_{k=0}^{s_2} \omega_k \frac{t^k}{k!}. 
\eea 
In Eq.\ (\ref{phaza}) the parameter $\Phi_0$ is the initial phase of the wave
form, $\bf{r}_{\rm SSB}$ is the vector joining the solar system
barycenter (SSB) with the detector, ${\bf n}_0$ is the constant
unit vector in the direction from the SSB to the source of 
the gravitational-wave signal.  
We assume that the gravitational-wave signal is almost monochromatic
around some angular frequency $\omega_0$ which we define as
instantaneous angular frequency evaluated at the SSB at $t=0$,
$\omega_k$ ($k=1,2,\ldots$) is the $k$th time derivative of the
instantaneous angular frequency at the SSB evaluated at $t=0$.  To
obtain formulas (\ref{phaza}) we model the frequency of the signal
in the rest frame of the neutron star by a Taylor series.  For the
detailed derivation of the phase model (\ref{phaza}) see Sec.\
II~B and Appendix A of Ref.\ \cite{JKS98}.  The integers $s_1$ and
$s_2$ are the number of spin downs to be included in the two
components of the phase. We need to include enough spin downs so
that we have a sufficiently accurate model of the signal to
extract it from the noise. 

The vector $\bf{r}_{\rm SSB}$ is the sum of two vectors:  the vector
$\bf{r}_{\rm ES}$ joining the solar system barycenter and the Earth barycenter
and vector $\bf{r}_{\rm E}$ joining the Earth barycenter and the detector.  The
vector $\bf{r}_{\rm ES}$ can be obtained form the JPL Planetary and Lunar
Ephemerides:  ``DE405/LE405''.  They are available via the Internet at anonymous
ftp:  \verb|navigator.jpl.nasa.gov|, the directory:  \verb|ephem/export|.
Whereas the vector $\bf{r}_{\rm E}$ can be accurately computed using the
International Earth Rotation Service (IERS) tables available at
\verb|http://hpiers.obspm.fr/iers/eop/eopc04|.  The codes to read the above
files and to calculate the vector $\bf{r}_{\rm SSB}$ are described in Ref.
\cite{puls4}.

It is convenient to write the response of the gravitational-wave
detector given above in the following form 
\be
\label{sig}
\fl h(t) = A_1 a(t)\cos\Phi(t) + A_2 b(t)\cos\Phi(t)
+ A_3 a(t)\sin\Phi(t) + A_4 b(t)\sin\Phi(t),
\ee
where $A_{i}$ are four constant amplitudes, the functions $a$ and $b$ 
are given by Eqs.\ (\ref{abb}),
and $\Phi$ is the phase given by Eq.\ (\ref{phaza}). The
modulation amplitudes $a$ and $b$ depend on the right ascension
$\alpha$ and the declination $\delta$ of the source (they also
depend on the angles $\phi$ and $\gamma$).  The phase $\Phi$
depends on the frequency $\omega_0$, $s$ spin-down parameters
$\omega_k$ $(k=1,\ldots,s)$, and on the angles $\alpha$, $\delta$.
We call the parameters $\omega_0$, $\omega_k$, $\alpha$, $\delta$ the
{\em intrinsic parameters} and the remaining ones the {\em extrinsic
parameters}. Moreover the phase $\Phi$ depends on the
latitude $\phi$ of the detector.  The whole signal $h$ depends on
$s+5$ unknown parameters:  $h_{0+}$, $h_{0\times}$, $\alpha$,
$\delta$, $\omega_0$, $\omega_k$.

\subsection{Optimal data analysis method}
\label{Sec:OPT} 

We assume that the noise $n$ in the detector is an
additive, stationary, Gaussian, and zero-mean random
process. Then the data $x$ (when the signal $h$ is present) can be
written as 
\be 
x(t) = n(t) + h(t). 
\ee 
The log likelihood function
has the form \be \label{loglr} \log\Lambda = (x|h) -
\frac{1}{2}(h|h), \ee where the scalar product
$(\,\cdot\,|\,\cdot\,)$ is defined by \be \label{SP} (x|y) := 4\,
\Re \int^{\infty}_{0} \frac{\tilde{x}(f)\tilde{y}^{*}(f)}{S_h(f)}
df. \ee 
In Eq.\ (\ref{SP}) $\Re$ denotes the real part, tilde denotes Fourier transform,
asterisk is complex conjugation, and $S_h$ is the {\em
one-sided} spectral density of the detector's noise. Assuming that
over the bandwidth of the signal $h$ the spectral density $S_h(f)$
is nearly constant and equal to $S_h(f_0)$, where $f_0$ is the
frequency of the signal measured at the SSB at $t=0$, the log
likelihood ratio from Eq.\ (\ref{loglr}) can be approximated by 
\be
\label{loglr3} \ln\Lambda \approx \frac{2T_o}{S_h(f_0)} \left(
\tav{xh}-\frac{1}{2}\tav{h^2} \right). 
\ee 
where $\tav{\cdot}$ denotes time averaging over the observational
interval $[0,T_o]$:
\be 
\tav{x} := \frac{1}{T_o} \int^{T_o}_{0} x(t)\,dt. 
\ee 

The signal $h$ depends linearly on four amplitudes $A_i$.  
The equations for the maximum likelihood (ML) estimators $\widehat{A}_{i}$ 
of the amplitudes $A_i$ are
given by \be \label{ampest} \frac{\partial\ln\Lambda}{\partial
A_{i}} = 0,\quad i=1,\ldots,4. \ee One can easily find the
explicit analytic solution to Eqs.\ (\ref{ampest}). To simplify
formulas we assume that {\em the observation time $T_o$ is an
integer multiple of one sidereal day}, i.e.,
$T_o=n(2\pi/\Omega_r)$ for some positive integer $n$. Then the
time average of the product of the functions $a$ and $b$ vanishes,
$\tav{ab}=0$, and the analytic formulas for the ML estimators of
the amplitudes are given by 
\be 
\label{amle0} 
\widehat{A}_1 \approx 2 \frac{\tav{x h_1}}{\tav{a^2}}, \hspace{3mm} 
\widehat{A}_2 \approx 2 \frac{\tav{x h_2}}{\tav{b^2}}, \hspace{3mm}
\widehat{A}_3 \approx 2 \frac{\tav{x h_3}}{\tav{a^2}}, \hspace{3mm}
\widehat{A}_4 \approx 2 \frac{\tav{x h_4}}{\tav{b^2}}.
\ee
The reduced log likelihood function $\F$ is the log likelihood function 
where amplitude parameters $A_i$ were replaced by their estimators 
$\widehat{A}_i$. 
By virtue of Eqs.\ (\ref{amle0}) from Eq.\ (\ref{loglr3}) one gets 
\be
\label{OS} \F \approx \frac{2}{S_h(f_0) T_o} \left(
\frac{|F_a|^2}{\tav{a^2}} + \frac{|F_b|^2}{\tav{b^2}} \right), 
\ee
where
\bea 
\label{Fab} 
F_{a} &:=& \int^{T_o}_0 x(t)\, a(t) \nonumber
\exp[-i\Phi(t)]\,dt,
\\[2ex]
F_{b} &:=& \int^{T_o}_0 x(t)\, b(t) \exp[-i\Phi(t)]\,dt.
\eea

The ML estimators of the signal parameters are obtained in two
steps.  Firstly, the estimators of the frequency, the spin-down parameters, and
the angles $\alpha$ and $\delta$ are obtained by maximizing the functional
$\F$ with respect to these parameters.  Secondly, the estimators of
the amplitudes $A_{i}$ are calculated from the analytic formulas (\ref{amle0})
with the correlations $\tav{xh_i}$ evaluated for the values of the parameters
obtained in the first step.

In order to calculate the optimum statistics $\F$ efficiently we introduce
an approximation to the phase of the signal that is valid
for observation times short comparing to the period of 1 year.
The phase of the gravitational-wave signal contains terms arising
from the motion of the detector with respect to the SSB.  These
terms consist of two contributions, one which comes from the
motion of the Earth barycenter with respect to the SSB, and the
other which is due to the diurnal motion of the detector with
respect to the Earth barycenter.  The first contribution has a
period of one year and thus for shorter observation times can be
well approximated by a few terms of the Taylor expansion.  The
second term has a period of 1 sidereal day and to a very good
accuracy can be approximated by a circular motion.  We find
that for two days of observation time that we used in the analysis
of the EXPLORER data presented in Section \ref{Sec:EXPs} we can 
truncate the expansion at terms that are quadratic in time.
Moreover in the Taylor expansion of the frequency parameter
it is sufficient to include terms up to the first spin down.
We thus propose the following {\em approximate} simple model of the phase
of the gravitational-wave signal: 
\be 
\label{Eq:Phs} 
\Psi_s(t) = 
p + p_0\, t + p_1\, t^2 +  A \cos(\Omega_r t) + B \sin(\Omega_r t), 
\ee 
where $\Omega_r$ is the rotational angular velocity of the Earth.
The characteristic feature of the above approximation is that the phase
is a linear function of the parameters of the signal.
The parameters $A$ and $B$ can be related
to the right ascension $\alpha$ and the declination $\delta$ of
the gravitational-wave source through the equations
\bea 
\label{AB}
A &=& \frac{\omega_0 r}{c} \cos\delta \cos(\alpha - \phi_r),
\\[2ex]
B &=& \frac{\omega_0 r}{c} \cos\delta \sin(\alpha - \phi_r),
\eea 
where $\omega_0$ is the angular frequency of the
gravitational-wave signal and $r$ is the equatorial component
of the detector's radius vector.
The parameters $p$, $p_0$, and $p_1$ contain contributions both form  
the intrinsic evolution of the gravitational-wave source
and the modulation of the signal due to the motion of the Earth 
around the Sun.
With this approximation the integrals given by Eqs.\ (\ref{Fab}) and needed to 
compute $\F$ become Fourier transforms and they can be efficiently
calculated using the FFT algorithm. Thus the evaluation of 
$\F$ consists of correlation of the data with two linear filters 
depending on parameters $p_1, A, B$ followed by FFTs. In Ref.\ \cite{puls4}
we have verified that for the case of our search the linear approximation 
to the phase does not lead to the loss of signal-to-noise ratio 
of more than 5\%.

In order to identify potential gravitational-wave signals we apply a two step
procedure consisting of a {\em coarse search} followed by a {\em fine search}.
The coarse search consists of evaluation of $\F$ on a discrete grid constructed
in such a way that the loss of the signal-to-noise is minimized and comparison
of the obtained values of $\F$ with a predefined threshold $\F_o$.  The
parameters of the nodes of the grid for which the threshold is crossed are
registered as potential gravitational-wave signals.  The threshold is calculated
from a chosen false alarm probability which is defined as the probability that
$\F$ crosses the threshold when no signal is present and the data is only noise.
The fine search consists of finding a local maximum of $\F$ using a numerical
implementation of the Nelder-Mead algorithm, where coordinates of the starting
point of the maximization procedure are the parameter values of the coarse
search.

To calculate the false alarm probability as a function of a threshold and to
construct a grid in the parameter space we introduce yet another approximation
of our signal.  Namely we use a signal with a constant amplitude and the phase
given by Eq.\ (\ref{Eq:Phs}).  In paper \cite{JK99} we have shown that the
Fisher matrix for the exact model with amplitude modulations given by Eq.\
(\ref{sig}) can be accurately reproduced by the Fisher matrix of a constant
amplitude model.  As the calculations of the false alarm probability and
construction of a grid in the parameter space are based on the Fisher matrix we
expect that the constant amplitude model is a good approximation for the purpose
of the above calculations.  We stress that in the search of real data we used
the full model with amplitude modulations.

We shall now summarize the statistical properties of the
functional $\F$.  We first assume that the phase parameters are
known and consequently that $\F$ is only a function of the random data
$x$.  When the
signal is absent $2\F$ has a $\chi^2$ distribution with four
degrees of freedom and when the signal is present it has a
noncentral $\chi^2$ distribution with four degrees of freedom and
noncentrality parameter equal to the \emph{optimal signal-to-noise
ratio} $d$ defined as
\be
\label{snr}
d := \sqrt{(h|h)}.
\ee
Consequently the probability density functions $p_0$ and $p_1$
when respectively the signal is absent and present are given by
\bea 
\label{p0} 
p_0(\F) &=& \F \exp(-\F),
\\[2ex]
\label{p1} p_1(d,\F) &=& \frac{\sqrt{2\F}}{d}
I_1\Big(d\sqrt{2\F}\Big) \exp\left(-\F-\frac{1}{2}d^2\right), 
\eea
where $I_1$ is the modified Bessel function of the first kind and
order 1. The false alarm probability $P_F$ is the probability that
$\F$ exceeds a certain threshold $\Fo$ when there is no signal. In
our case we have
\be
\label{PF}
P_F(\Fo) := \int_{\Fo}^\infty
p_0(\F)\, d\F = (1 + \Fo) \exp(-\Fo).
\ee
The probability of
detection $P_D$ is the probability that $\F$ exceeds the threshold
$\Fo$ when the signal-to-noise ratio is equal to $d$: 
\be
\label{PD} 
P_D(d,\Fo) := \int_{\Fo}^{\infty} p_1(d,\F)\, d\F. 
\ee
The integral in the above formula cannot be evaluated in terms of
known special functions. 

We see that when the noise in the
detector is Gaussian and the phase parameters are known the
probability of detection of the signal depends on a single
quantity: the optimal signal-to-noise ratio $d$. When the phase
parameters are unknown the optimal statistics $\F$ depends not
only on the random data $x$ but also on the phase parameters that
we shall denote by $\bf{\xi}$. Such an object is called in the
theory of stochastic processes a {\em random field}.  Let us
consider the correlation function of the random field 
\be
\label{covdef} 
C({\bf \xi},{\bf \xi}') := E \left\{
[\F({\bf \xi})-m({\bf \xi})] [\F({\bf \xi}')-m({\bf \xi}')] \right\},
\ee 
where 
\be 
m({\bf\xi}) := E \left\{\F({\bf\xi})\right\}. 
\ee
For the case of the constant amplitude, linear phase model we have  
\be
\label{gr2}
m({\bf\xi}) = 1,
\ee
and
\bea
\label{gr3}
C({\bf\xi},{\bf\xi}') & = C({\bf\xi} - {\bf\xi}')
\nonumber\\
& = \tav{\cos[\Delta p_0 t + \Delta p_1 t^2 +
\Delta A \cos(\Omega_r t) + \Delta B \sin(\Omega_r t)]}^2
\nonumber\\
& \quad + \tav{\sin[\Delta p_0 \, t + \Delta p_1 \, t^2 +
\Delta A \, \cos(\Omega_r t) + \Delta B \, \sin(\Omega_r t)]}^2,
\eea
where $\Delta$ denotes the difference between the parameter values.
Thus the correlation function $C$ depends only on the difference
$\bf{\xi}-\bf{\xi}'$ and not on the values of the parameters
themselves. In this case the random field $\F$ is called {\em
homogeneous}. For such fields we have developed \cite{JK00} a
method to calculate the false alarm probability that we only summarize here.
The main idea is to divide the space of the intrinsic parameters $\bf{\xi}$ 
into {\em elementary cells}.  The size of the cell is determined by the
\emph{characteristic correlation hypersurface} of the random field
$\F$.  The correlation hypersurface is defined by the condition
that at this hypersurface the correlation $C$ equals to half its
maximum value.  Assuming that $C$ attains its maximum value
when $\bf{\xi}-\bf{\xi}'=$ 0 the equation of the characteristic
correlation hypersurface reads 
\be 
\label{gcov1} 
C({\bf\tau}) = \frac{1}{2} C(0), 
\ee 
where we have introduced $\bf{\tau}:=\bf{\xi}-\bf{\xi}'$. 
Let us expand the left-hand side of Eq.\ (\ref{gcov1}) 
around $\bf{\tau}=$ 0 up to terms of the second
order in $\bf{\tau}$. We arrive at the equation 
\be 
\label{gcov2}
\sum_{i,j=1}^M G_{ij} \tau_i\tau_j = 1, 
\ee 
where $M$ is the dimension of the intrinsic parameter space 
and the matrix $G$ is defined as follows 
\be 
\label{gmatrix} 
G_{ij} := -\frac{1}{C(0)}
\frac{\pa^2{C(\bf{\tau})}}{\pa{\tau_i}\partial{\tau_j}}
\Bigg\vert_{\bf{\tau}=0}. 
\ee 
Equation (\ref{gcov2}) defines the boundary of an $M$-dimensional 
hyperellipsoid which we call the \emph{correlation hyperellipsoid}. 
The $M$-dimensional hypervolume $V_{cell}$ of the hyperellipsoid
defined by Eq.\ (\ref{gcov2}) equals 
\be 
\label{vc} 
V_{cell} =
\frac{\pi^{M/2}}{\Gamma(M/2+1)\sqrt{\det G}}, 
\ee 
where $\Gamma$ denotes the Gamma function.
We estimate the number $N_c$ of elementary cells by dividing the total
Euclidean volume $V_{total}$ of the parameter space by the volume
$V_{cell}$ of the correlation hyperellipsoid, i.e.\ we have
\be
\label{NT}
N_c = \frac{V_{total}}{V_{cell}}.
\ee
We approximate the probability distribution of $\F(\bf{\xi})$ in
each cell by the probability distribution $p_0(\F)$ when the
parameters are known [in our case it is the one given by Eq.\ (\ref{p0})].
The values of the statistics $\F$ in each cell can be considered
as independent random variables.  The probability that $\F$ does
not exceed the threshold $\Fo$ in a given cell is $1-P_F(\Fo)$,
where $P_F(\Fo)$ is given by Eq.\ (\ref{PF}).  Consequently the
probability that $\F$ does not exceed the threshold $\Fo$ in
\emph{all} the $N_c$ cells is $[1-P_F(\Fo)]^{N_c}$.  The
probability $P^T_F$ that $\F$ exceeds $\Fo$ in \emph{one or more}
cells is thus given by 
\be 
\label{FP} 
P^T_F(\Fo) = 1 - [1 - P_F(\Fo)]^{N_c}. 
\ee 
This is the false alarm probability when the phase parameters are unknown. 

The basic quantity to consider in the construction of the grid
of templates in the parameter space is the expectation value 
$E_1\left\{\F\right\}$ 
of the statistics when the signal is present. 
For the constant amplitude linear phase model we have that
\begin{equation}
E_1\left\{\F(\xi)\right\} = 1 + \frac{1}{2}d^2 C(\xi),
\end{equation}
where $C$ is the correlation function of the random field
introduced above and $d$ is the signal-to-noise ratio of the signal
present in the data. We choose the grid of templates in such a way
that the correlation between any potential signal present in the
data and the nearest point of the grid never falls below a certain
value. In the case of the approximate model of the signal that we
use the grid is unform and consists of regular polygons in
the space parametrized by $p_0, p_1, A, B$. The construction of 
the grid is described in detail in Section VIIA of \cite{ROGc00}.
For the grid that we used in the search of the EXPLORER data 
the correlation function $C$ for any signal present
in the data was greater than 0.77.

\section{An all-sky search of the EXPLORER data}
\label{Sec:EXPs}

We have implemented the theoretical tools presented in Section \ref{Sec:Met} 
and we have performed an all-sky search for continuous 
sources of gravitational waves in the data of the resonant bar detector
EXPLORER\footnote[1]{The EXPLORER detector is operated by the ROG
collaboration located in Italian Istituto Nazionale di Fisica
Nucleare (INFN); see \tt http://www.roma1.infn.it/rog/explorer/explorer.html.}
\cite{exp}. The EXPLORER detector has collected many years of data with a
high duty cycle (e.g.\ in 1991 the duty cycle was 75\%). The
EXPLORER detector was most sensitive over certain two narrow
bandwidths (called minus and plus modes) of about 1~Hz wide at two
frequencies around 1~kHz. To make the search computationally manageable 
we analyzed two days of data in the narrow band were the detector had
the best sensitivity.  By narrowing the bandwidth of the search we
can shorten the length of the data to be analyzed as we need to
sample the data at only twice the bandwidth.  
For the sake of the FFT algorithm it is best to keep
the length of the data to be a power of $2$. Consequently we have
chosen the number of data points to analyze to be $N=2^{18}$. Thus
for $T_o=2$ days of observation time the bandwidth $\Delta\nu$
was $\Delta\nu=N/(2T_o)\sim0.76$~Hz. We have also chosen
to analyze the data for the plus mode. As a result we searched 
the bandwidth from 921.00 Hz to 921.76 Hz. 
We have used the filters with the phase linear in the
parameters given by Eq. (\ref{Eq:Phs}). In the filters we have 
included the amplitude modulation. The number of cells $N_c$
calculated from Eq.\ (\ref{NT}) was around $1.6 \times 10^{12}$.
Consequently from Eq.\ (\ref{FP}) the threshold signal-to-noise ratio 
for 1\% false alarm probability was equal to $8.3$. In the search 
that we have performed we have used a lower threshold signal-to-noise 
of $6.7$.
The aim of lowering the threshold was to make up for the loss of
the signal-to-noise ratio due the discreteness of the grid of templates
and due to the use of filters that only approximately matched the true signal.
The number of points in the grid over which we had to calculate
the statistics $\F$ turned out to be $183\,064\,440$. This number
involved $63\,830$ positions in the sky and $2\,868$ spin
down values for each sky position. 
We have carried out the search on a network of PCs and workstations. 
We had around two dozens of processors at our disposal. The data analysis
started in September 2001 and was completed in November 2002.

\begin{figure}[t]
\centering
\includegraphics[width=8cm]{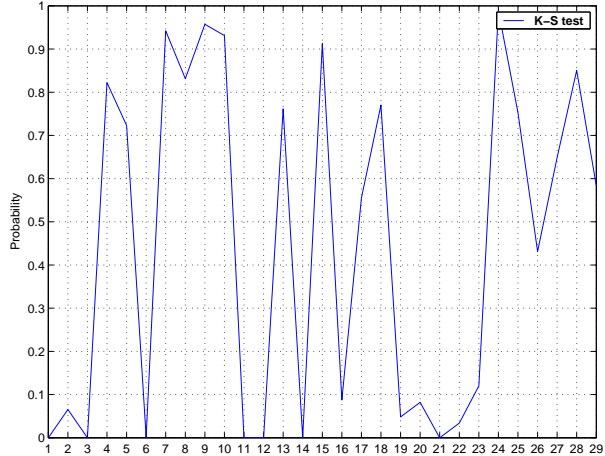}
\caption{Quality of the EXPLORER data. The probability values of the KS 
statistics are given for each data segment from the $13$ day data run.}
\label{fig2}
\end{figure}

\begin{figure}[t]
\centering
\includegraphics[width=12cm]{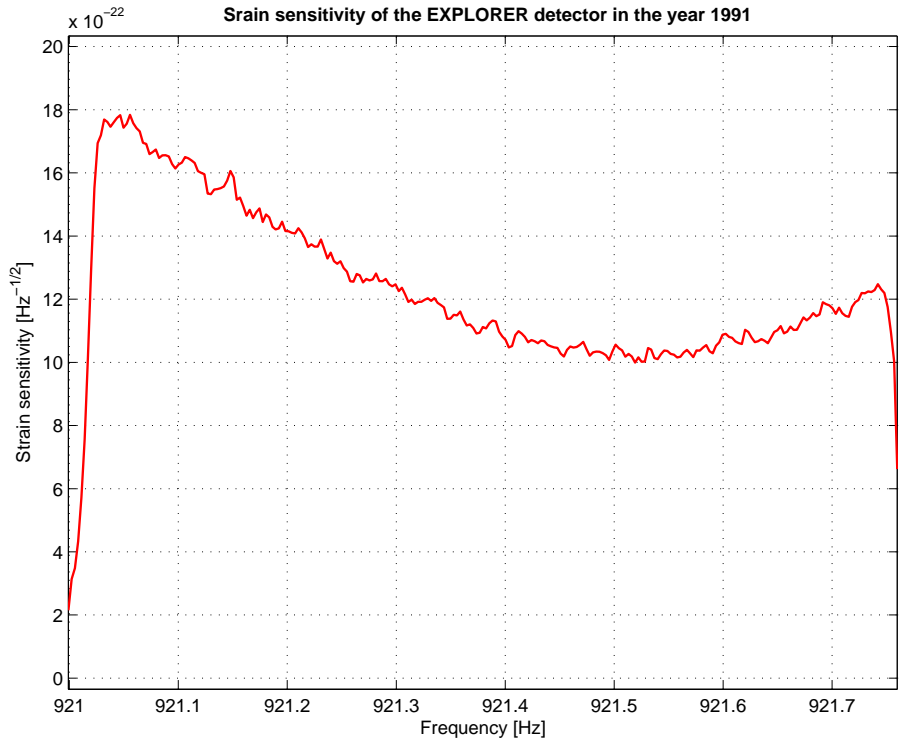}
\caption{Two-sided spectral density of the EXPLORER data.}
\label{fig1}
\end{figure}

The two-day stretch of data that we analyzed was taken from 
a larger set of $13$ days of data taken by the EXPLORER detector
in November 1991. We have chosen the 2-day stretch
of data on the basis of conformity of the data to the Gaussian random process.
We have divided the data into $2^{16}$ points sections corresponding 
to around $11$ hours of data. For each stretch we
have performed the Kolmogorov-Smirnov (KS) test.
The results of the KS test are presented in Figure \ref{fig2}.
For the case of KS test the higher the probability the more Gaussian the data 
are. {From} the above test we conclude that large parts of data 
are approximately Gaussian. On the basis of the above analysis
we have chosen the two day data stretch to begin at the first sample
of the 7th data stretch corresponding to Modified Julian Date
of $48580.7909$. In Figure \ref{fig1} we have presented 
the spectral density of the two day stretch of data that we analyzed.
We see that the minimum spectral density was close to 
$10^{-21}/\sqrt{\mbox{Hz}}$.

\begin{figure}[t]
\centering
\includegraphics[width=8cm]{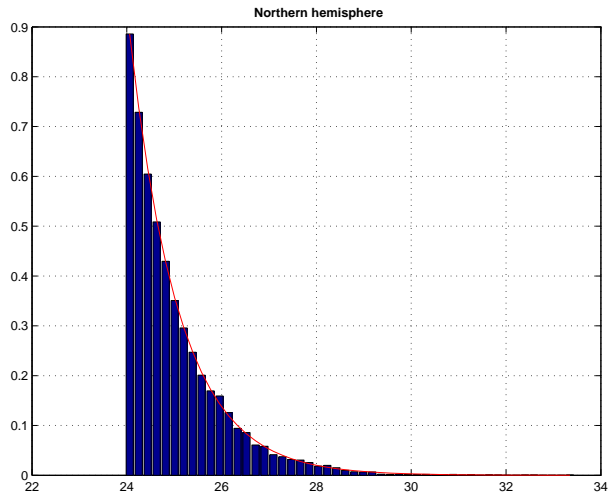}
\caption{\label{fig4}
Histogram of candidates: Northern hemisphere. On the $x$-axis the values of the 
statistic $\F$ are given. The solid line corresponds to the theoretical $\chi^2$ 
distribution with 4 degrees of freedom.}
\end{figure}

\begin{figure}[t]
\centering
\includegraphics[width=8cm]{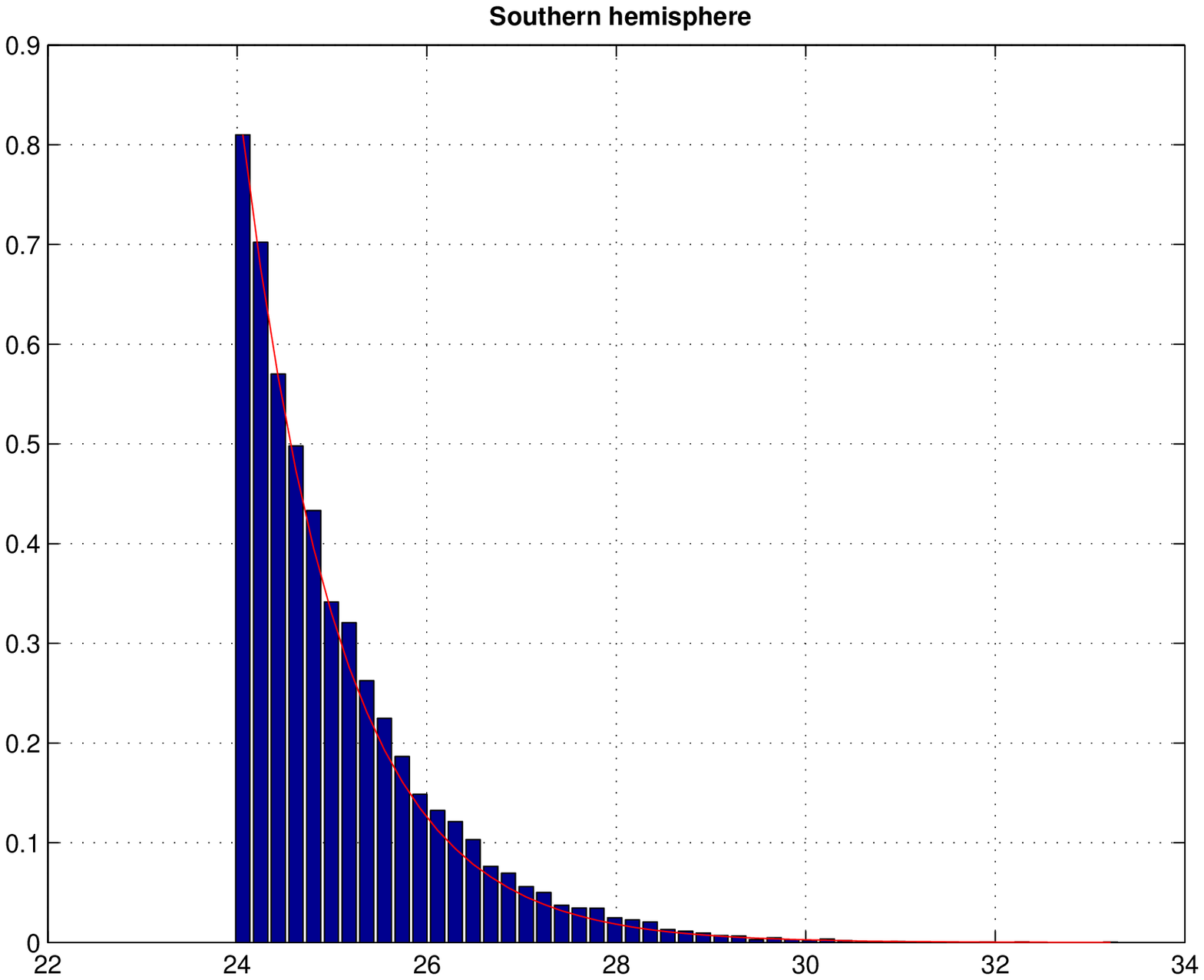}
\caption{\label{fig5}
Histogram of candidates: Southern hemisphere.}
\end{figure}

We have obtained $22\,295$ threshold crossings for the Northern hemisphere and
$44\,701$ for the Southern hemisphere.  We consider all candidates contained
within a single cell of the parameter space defined in Section \ref{Sec:OPT} as
dependent and we choose as an independent candidate the candidate that has the
highest signal-to-noise ratio within one cell.  Consequently we have obtained
$11\,703$ independent candidates for the Northern hemisphere and $18\,702$ for
the Southern hemisphere.  In Figures \ref{fig4} and \ref{fig5} we have plotted
the histograms of the values of the statistics $\F$ for the independent
candidates and we have compared it with the theoretical distribution for $\F$
when no signal is present in the data.  A good agreement with the theoretical
distribution is another indication of Gaussianity of the data.  It also reveals
that there are no obvious populations of the continuous gravitational-wave
sources at the level of sensitivity of our search.  In the search no event has
crossed our $99\%$ confidence threshold of $8.3$.  The strongest signal obtained
by the coarse search had the signal-to-noise ratio of $7.9$ and the fine search
increased that value to $8.2$.

As we do not have a detection of a gravitational-wave signal we can make a
statement about the upper bound for the gravitational-wave amplitude.  To do
this we take our strongest candidate of signal-to-noise ratio $d_o$ and we
suppose that it resulted from a gravitational-wave signal.  Then, using formula
(\ref{PD}), we calculate the signal-to-noise $d_{ul}$ of the gravitational-wave
signal so that there is $1\%$ probability that it crosses the threshold $\F_o$
corresponding to $d_o$, where $\F_o=2+\frac{1}{2}d^2_o$.  The $d_{ul}$ is the
desired $99\%$ confidence upper bound.  For $d_o = 8.2$, which corresponds to
the signal-to-noise ratio of our strongest candidate, we find that $d_{ul} =
5.9$.  For the EXPLORER detector this corresponds to the dimensionless amplitude
of the gravitational-wave signal equal to $2\times10^{-23}$.  Thus we have the
following conclusion:

{\em In the frequency band from 921.00 Hz to 921.76 Hz and for signals coming
from any sky direction the dimensionless amplitude of the gravitational-wave
signal from a continuous source is less than ${\bf 2\times10^{-23}}$ with $99\%$
confidence.}

\section*{Acknowledgments}

The work of P.\ Jaranowski, A.\ Kr\'olak, K.\ M.\ Borkowski, and M.\ Pietka was
supported in part by the KBN (Polish State Committee for Scientific Research)
Grant No.\ 2P03B 094 17.  The part of A.\ Kr\'olak research described in this
paper was performed while he held an NRC-NASA Resident Research Associateship at
the Jet Propulsion Laboratory, California Institute of Technology, under
contract with the National Aeronautics and Space Administration.  We would also
like to thank Interdisciplinary Center for Mathematical and Computational
Modeling of Warsaw University for computing time.

\end{document}